\documentclass[english,prd,twocolumn,nofootinbib]{revtex4}
\usepackage[T1]{fontenc}
\usepackage[latin1]{inputenc}
\usepackage{babel}
\usepackage{graphics}
\input epsf


\begin{document}

\title{Kink interactions in $SU(N)\times Z_2$}

\author{
Levon Pogosian}
\affiliation{
Theoretical Physics, The Blackett Laboratory, Imperial College,
Prince Consort Road, London SW7 2BZ, United Kingdom. }

\begin{abstract}
There are $N-1$ classes of kink solutions in $SU(N)\times Z_2$. We
show how interactions between various kinks depend on the classes
of individual kinks as well as on their orientations with respect
to each other in the internal space. In particular, we find that
the attractive or repulsive nature of the interaction depends on
the trace of the product of charges of the two kinks.
We calculate the interaction potential for all combinations of kinks and antikinks in
$SU(5)\times Z_2$ and study their collisions. The outcome of
kink-antikink collisions, as expected from previous studies,
is sensitive to their initial relative velocity. We find
that heavier kinks tend to break up into lighter ones, while
interactions between the lightest kinks and antikinks in this model
can be repulsive as well as attractive.
\end{abstract}

\maketitle

\section{\label{introduction}Introduction}

Topological defects are observed in condensed matter systems and
may have been formed during phase transitions\footnote{Here,
the term ``phase transition'' includes continuous transitions called crossovers.} at
early stages in the history of the universe \cite{VilShel}. If observed, they
would provide invaluable information about the universe when it was a tiny
fraction of a second old. If not observed, topological defects still play
an important role by placing constraints on particle physics models and cosmology.
The formation and scaling of a network of defects strongly depends on how they
interact among themselves. This, in turn, will affect the type and the strength of
restrictions that observations (or the lack thereof) can impose on the underlying
model. Another context, in which interactions between defects are important, is
the possible connection between elementary particles and solitonic solutions of
classical field equations \cite{Raj}.

In addition to the more general reasons given above, understanding
how the $SU(N)\times Z_2$ domain walls interact could be
important in the light of the correspondence, found by
Vachaspati \cite{Vac96}, between the spectrum of $SU(5)$ monopoles
and the spectrum of one family of fermions in the Standard Model.
Interactions between $SU(N)\times Z_2$ kinks may also be
relevant to the solution of the monopole over-abundance problem
based on sweeping monopoles with domain walls as proposed in
\cite{DvaLiuVac98}.

Previous work on interactions between kinks has mainly concentrated on
the sine-Gordon and the $\phi^4$ models in (1+1) dimensions
\cite{old1,old2,old3,old4,old5,Moshir81,Sugiyama79,Campbell_etal83}.
Even a relatively simple system of a $\phi^4$ kink interacting with
an antikink can have rather non-trivial dynamics, which is one of the reasons
why so many researchers have worked on this problem in the past.
The force between kinks and antikinks of the $\phi^4$ model is always attractive.
The outcome of their collision can be one of the three types: they can annihilate,
they can scatter off each other or they can
form an intermediate bound state before ultimately separating or annihilating.
The general tendency is that at low collision velocities kinks tend to annihilate
and at higher velocities they scatter. However, the dependence of the outcome
on the incident velocity is rather non-linear, as was found in
\cite{old1,old2,old3,old4,old5,Moshir81} and investigated in detail
in \cite{Campbell_etal83}. Namely, it was found that, over a relatively small range
of initial velocities, intervals of initial velocity for which kink and antikink
capture each other alternate with regions for which the interaction concludes with escape
to infinite separations. In \cite{Campbell_etal83}, this alternation phenomenon was
attributed to a nonlinear resonance between the orbital frequency of the bound
kink-antikink pair and the frequency of characteristic small oscillations of the
field localized at the moving kink and antikink centers.

In this work, when discussing interactions between kinks,
we will aim to concentrate on issues that are unique to $SU(N)\times Z_2$
and will refer to earlier work when a problem can be reduced to that of
kinks in the $\phi^4$ model. In particular, we will show that in $SU(N)\times Z_2$
kinks can repel as well as attract.

This paper is organized as follows. In Sec. \ref{kinks} we give a brief overview
of kink solutions in $SU(N)\times Z_2$. In Sec. \ref{interactionsN} we develop a
framework in which $SU(N)\times Z_2$ kink interactions can be discussed and show
a simple way of determining whether a given pair of kinks will attract or repel.
In Sections \ref{interactions5} and \ref{collisions} we study the kink-antikink
interactions in $SU(5)\times Z_2$. Results are summarized in Sec. \ref{conclude}

\section{Kinks in $SU(N)\times Z_2$}
\label{kinks}

Consider a (1+1)-dimensional model of a scalar field $\Phi$
transforming in the adjoint representation of $SU(N)$, with $N$
taken to be odd and with the additional $Z_2$ symmetry that takes
$\Phi$ to $-\Phi$. The Lagrangian is
\begin{equation}
L = {\rm Tr} (\partial_\mu \Phi )^2 - V(\Phi ) \ ,
\label{lagrangian_N}
\end{equation}
where $V(\Phi )$ is such that $\Phi$ has an expectation value $\Phi_V $
that can be chosen to be
\begin{equation}
\Phi_V = \eta \sqrt{2 \over {N(N^2-1)}}
                  \pmatrix{n{\bf 1}_{n+1}&{\bf 0}\cr
                      {\bf 0}&-(n+1){\bf 1}_n\cr} \ ,
\label{phiV}
\end{equation}
where ${\bf 1}_p$ is the $p\times p$ identity matrix, $N\equiv 2n+1$ and $\eta$ is
an energy scale determined by the minima of the potential $V$.
Such an expectation value spontaneously breaks the symmetry down
to:
\begin{equation}
H = [SU(n+1)\times SU(n)\times U(1)]/[Z_{n+1}\times Z_{n}] \label{unbrokensymm} \, .
\end{equation}

Various types of kink solutions in this model are defined by
choices of the boundary conditions at $x=-\infty $ and
$x=+\infty$, where $x$ is the space coordinate. It was proved in
\cite{PogVac01} that for a kink solution to exist one must
necessarily have $[\Phi(-\infty),\Phi(+\infty)]=0$. This allows
one to list all the possible boundary conditions (up to gauge
rotations) that can lead to kink solutions. We can fix $\Phi_L
\equiv \Phi(-\infty)=\Phi_V$ given in Eq. (\ref{phiV}). Then we
can have
$$
\Phi_R \equiv \Phi(+\infty)=
 \epsilon_T \eta \sqrt{2 \over {N(N^2-1)}} \times
$$
\begin{equation}
{\rm diag} ( n{\bf 1}_{n+1-q}, -(n+1){\bf 1}_{q},
                         n{\bf 1}_{q}, -(n+1){\bf 1}_{n-q}) \ ,
\label{phi+choices_N}
\end{equation}
where we have introduced a parameter $\epsilon_T =\pm 1$ and
another $q=0,...,n$. The label $\epsilon_T$ is $+1$ when the
boundary conditions are topologically trivial and is $-1$ when
they are topologically non-trivial. $q$ tells us how many diagonal
entries of $\Phi_L$ have been permuted in $\Phi_R$. The case $q=0$
is when $\Phi_R = \epsilon_T \Phi_L$.
As was suggested in \cite{PogVac01}, the lowest energy stable topological ($\epsilon_T=-1$)
kink solution corresponds to $q=n$. Topological $q=n$ kinks were
studied in detail in \cite{Vac01,PogVac00}.

The most general ansatz for the kink solution was found in \cite{PogVac01}
and can be written as:
\begin{equation}
\Phi_k = F_+ (x) {\bf M_+} + F_- (x) {\bf M_-} + g(x) {\bf M}\ ,
\label{kinkexplicit2}
\end{equation}
where
\begin{equation}
{\bf M}_+ =  {{\Phi_R + \Phi_L}\over {2}} \ , \ \ {\bf M}_- =
{{\Phi_R - \Phi_L}\over {2}} \label{M+M-} \, .
\end{equation}
and, for $q\ne 0$ and $q\ne n$,
\begin{eqnarray}
{\bf M} = \mu \, {\rm diag} ( q(n-q){\bf 1}_{n+1-q},
   -(n-q)(n+1-q){\bf 1}_q , \nonumber \\ -(n-q)(n+1-q){\bf 1}_q,
q(n+1-q){\bf 1}_{n-q} ) \ \ \ \label{Mresult}
\end{eqnarray}
with
\begin{equation}
\mu = \eta [ 2q(n-q)(n+1-q)\{ 2n(n+1-q)-q\} ]^{-1/2} \ .
\label{muvalue}
\end{equation}
For $q=0$ or for $q=n$, the matrix ${\bf M}$ is zero. The boundary
conditions for $F_\pm$ and $g(x)$ are:
\begin{equation}
F_- (\pm \infty ) = \pm 1 \ ,
\ \ F_+ ( \pm \infty ) = 1 \ , \ \  g (\pm \infty ) = 0 \ .
\label{Fpmbc}
\end{equation}

One can define the charge of the kinks as
\cite{Raj}
\begin{equation}
Q \equiv {1 \over \eta}\left(\Phi_k (+\infty) - \Phi_k (-\infty) \right),
\label{charge}
\end{equation}
which corresponds to a current
\begin{equation}
j^\mu \equiv {1 \over \eta} \varepsilon^{\mu \nu} \partial_\nu \Phi \, ,
\end{equation}
where $\mu,\nu=0,1$, and $\varepsilon^{\mu \nu}$ is the antisymmetric tensor.
The definition (\ref{charge}) of $Q$ can be used for non-topological kinks as well as
topological ones.

\section{Kink interactions in $SU(N)\times Z_2$}
\label{interactionsN}

Consider two kinks, $K^{(1)}$ and $K^{(2)}$, separated by a distance
which is larger than their core sizes.
The classes to which the two kinks belong, as well as the global topology of
the two-kink configuration, are determined by the
choices of the three vacua:
\begin{itemize}
\item $\Phi_-$ at $x=-\infty$,
\item $\Phi_0$ in between the two kinks,
\item $\Phi_+$ at $x=+\infty$ .
\end{itemize}

Let indices ($\epsilon_T^{(1)},q^{(1)}$) describe the kink between
$\Phi_-$ and $\Phi_0$, where $q^{(1)}$ denotes the kink class as
defined in Sec. \ref{kinks}, and $\epsilon_T^{(1)}$ is $-1$ for
topological and $+1$ for non-topological kinks. Similarly, let
($\epsilon_T^{(2)},q^{(2)}$) denote the kink bounded by $\Phi_0$
and $\Phi_+$. Boundaries $\Phi_-$ and $\Phi_+$ of the two-kink
system can also be described using indices
($\epsilon_T^{(3)},q^{(3)}$) defined in a similar way . Topology
requires that $\epsilon_T^{(3)}=\epsilon_T^{(1)}
\epsilon_T^{(2)}$. Thus, the two-kink system can be described by
($q^{(1)},q^{(2)},q^{(3)},\epsilon_T^{(1)},\epsilon_T^{(2)}$).
This notation is invariant under global gauge rotations, since it
contains only information about how many diagonal entries were
permuted in each of the vacua $\Phi_-$, $\Phi_0$ and $\Phi_+$ with
respect to each other.

For given values of $q^{(1)}$ and $q^{(2)}$, not all values of
$q^{(3)}$ will generally be allowed. To determine the selection
procedure, let us start with $\Phi_0$, in which $q^{(1)}$ diagonal
entries of $\Phi_-$ were permuted. We want to know how many
diagonal entries of $\Phi_-$ can be permuted in $\Phi_+$ ({i.e.}
the value of $q^{(3)}$) given that $q^{(2)}$ diagonal entries of
$\Phi_0$ were permuted in $\Phi_+$. There are $n+1$ entries with
absolute values equal to $n$ in $\Phi_-$ (see Eq. (\ref{phiV})),
which we can denote by ${\cal A}$'s, and $n$ entries with absolute values
equal to $(n+1)$, which we will denote by ${\cal B}$'s. We will refer to
those ${\cal A}$'s and ${\cal B}$'s of $\Phi_0$, that were permuted to form
$\Phi_0$ out of $\Phi_-$, as ``changed'', and the ones that were
left untouched as ``unchanged''. Then, when permuting ${\cal A}$'s and
${\cal B}$'s of $\Phi_0$ to form $\Phi_+$ ($q^{(2)}$ permutations ), one
has the following options:
\begin{enumerate}
\item Permute a changed ${\cal A}$ with a changed ${\cal B}$. This operation will
decrease the value of $q^{(3)}$ by $1$, since it will restore the
original order of the given pair of ${\cal A}$ and ${\cal B}$ in $\Phi_-$. Let
$q^{(2)}_1$ denote the number of possible ways it can be done.
Since there are only $q^{(1)}$ changed ${\cal A}$'s, $q^{(2)}_1 \le {\rm
min}(q^{(1)},q^{(2)})$. Also, if there is a deficit of unchanged
${\cal A}$'s, one is forced to permute at least $(q^{(2)}-n+q^{(1)}-1)$
changed $A's$, which means that $q^{(2)}_1 \ge {\rm
max}(0,q^{(2)}-n+q^{(1)}-1)$.

\item Permute a changed ${\cal A}$ with an unchanged ${\cal B}$. This operation
does not affect the value of $q^{(3)}$. The number of possible
ways in which this can be done, denoted by $q^{(2)}_2$, is limited
by the number of available changed ${\cal A}$'s and unchanged ${\cal B}$'s:
$q^{(2)}_2 \le {\rm min}(q^{(1)}-q^{(2)}_1,n-q^{(1)})$.

\item Permute an unchanged ${\cal A}$ with a changed ${\cal B}$. This operation
also does not change the value of $q^{(3)}$. It can be done in
$q^{(2)}_3$ ways, limited by the number of available unchanged
${\cal A}$'s and changed ${\cal B}$'s: $q^{(2)}_3 \le {\rm
min}(n-q^{(1)}+1,q^{(1)}-q^{(2)}_1)$.

\item Permute an unchanged ${\cal A}$ with an unchanged ${\cal B}$. This operation
will increase the value of $q^{(3)}$ by $1$, and can be performed
in $q^{(2)}_4$ ways, limited by the number of available unchanged
${\cal A}$'s and unchanged ${\cal B}$'s: $q^{(2)}_4 \le {\rm
min}(n-q^{(1)}-q^{(2)}_3+1,n-q^{(1)}-q^{(2)}_2)$.
\end{enumerate}
In summary, independent of $\epsilon_T^{(1)}$ and
$\epsilon_T^{(2)}$, given $q^{(1)}$ and $q^{(2)}$, the set of
possible values of $q^{(3)}$ can be found using
\begin{equation}
q^{(3)} = q^{(1)} - q^{(2)}_1 + q^{(2)}_4 \, ,
\end{equation}
where we have defined integers $q^{(2)}_i$ ($i=1,2,3,4$) that can
take on all non-negative values allowed by the following selection
rules:
\begin{eqnarray}
\nonumber
q^{(2)}_1+q^{(2)}_2+q^{(2)}_3+q^{(2)}_3=q^{(2)} \ \ \ \ \ \ \, , \\
\nonumber
{\rm max}(0,q^{(2)}-n+q^{(1)}-1) \le q^{(2)}_1 \le {\rm min}(q^{(1)},q^{(2)}) \, ,\\
\nonumber
               0  \le q^{(2)}_2 \le {\rm min}(q^{(1)}-q^{(2)}_1,n-q^{(1)}) \ \ \ \ \, , \\
\nonumber
               0  \le q^{(2)}_3 \le {\rm min}(n-q^{(1)}+1,q^{(1)}-q^{(2)}_1) \ \ \ \  \, ,\\
               0  \le q^{(2)}_4 \le
               {\rm min}(n-q^{(1)}-q^{(2)}_3+1,n-q^{(1)}-q^{(2)}_2) \, .
\end{eqnarray}

Next we would like to determine whether a given pair of kinks will
attract or repel. We can use a procedure which is analogous to one
used by Manton \cite{Manton79} in the case of the $\phi^4$.
At large separations, the two-kink ansatz can be written as:
\begin{equation}
\Phi(x)=\Phi_k^{(1)}(x+a)+\Phi_k^{(2)}(x-a)-\Phi_0 \, ,
\end{equation}
where $\Phi_k^{(1)}(x)$ is the first kink solution with $\Phi_k^{(1)}(-\infty)=\Phi_-$,
$\Phi_k^{(2)}(x)$ is the second kink solution with $\Phi_k^{(2)}(+\infty)=\Phi_+$,
$\Phi_0 \equiv \Phi_k^{(1)}(+\infty)=\Phi_k^{(2)}(-\infty)$ is the vacuum
in the region separating the kinks and $a>0$ is the distance from the origin
to the centers of the kinks.
Using Eqns. (\ref{kinkexplicit2}) and (\ref{charge}),
we can write:
\begin{eqnarray}
\Phi(x) = F_+^{(1)}M_+^{(1)} + {\eta \over 2} F_-^{(1)}Q^{(1)}+ g^{(1)}M^{(1)}
\nonumber \\
+ F_+^{(2)}M_+^{(2)}+ {\eta \over 2} F_-^{(2)}Q^{(2)} + g^{(2)}M^{(2)} - \Phi_0 \, .
\label{2kinks}
\end{eqnarray}
In \cite{PogVac01} it was shown that functions $F_+^{(1,2)}$ and
$g^{(1,2)}$ can be treated as approximately constant for a
relatively wide range of parameters of a general quartic
potential. A simplified (and, therefore, only approximate) version of
the two-kink ansatz is given by:
\begin{eqnarray}
\Phi(x) &\approx& M^{(1)}_+ + {\eta \over 2} F^{(1)}_- Q^{(1)} \nonumber \\
&+& M^{(2)}_+ + {\eta \over 2} F^{(2)}_- Q^{(2)} - \Phi_0 \ ,
\label{easy2kinks}
\end{eqnarray}
where
\begin{eqnarray}
M^{(1)}_+ \equiv {\Phi_- + \Phi_0 \over 2 } \ , \ \
M^{(2)}_+ \equiv {\Phi_0 + \Phi_+ \over 2 } \ ,
\end{eqnarray}
and
\begin{equation}
F^{(1)}_- \approx {\rm tanh}[\sigma (x+a)] \ ,
\ F^{(2)}_- \approx {\rm tanh}[\sigma (x-a)] \ ,
\end{equation}
where $a \gg \sigma^{-1}$ and $\sigma^{-1}$ is the ``width'' of the wall.

The field energy-momentum tensor can be derived using the action principle and
is given by
\begin{equation}
T^{\mu \nu}= 2{\rm Tr}[\partial^{\mu} \Phi \partial^{\nu} \Phi] -
\eta^{\mu \nu} {\rm Tr}[\partial^{\sigma} \Phi \partial_{\sigma} \Phi] - \eta^{\mu \nu} V(\Phi) \ .
\end{equation}
Therefore, the momentum density in (1+1) dimensions is
\begin{equation}
{\cal P} \equiv T_{0}^{1}= - 2{\rm Tr}[\dot{\Phi} \Phi'] \ ,
\end{equation}
where $\dot{\Phi} \equiv \partial_t \Phi$ and $\Phi' \equiv \partial_x \Phi$.
One can define the momentum of a field configuration on the interval $x_1<x<x_2$ as
\begin{equation}
P = - \int_{x_1}^{x_2} 2{\rm Tr}[\dot{\Phi}\Phi']dx \ .
\end{equation}
To calculate the force between the kinks, let us consider the initial rate
of change of momentum of a (intially) static field configuration given by the
two-kink ansatz:
\begin{equation}
{dP \over dt} = - \int_{x_1}^{x_2}
2({\rm Tr}[\ddot{\Phi}\Phi']+{\rm Tr}[\dot{\Phi}\dot{\Phi}']) dx \ .
\end{equation}
One can use field equations of motion and integrate to obtain
\begin{equation}
{dP \over dt} = \Big[ - {\rm Tr}[\dot{\Phi}^2] -
{\rm Tr}[(\Phi')^2] + V(\Phi) \Big]_{x_1}^{x_2} \ .
\end{equation}
Let us choose $x_1 \ll -a$ and $ -a \ll x_2 \ll a$ ({\it e.~g.} $x_2=0$).
That is, we want to estimate the force on the first kink (the one at $x=-a$),
due to the second kink. Let us define
\begin{equation}
f \equiv {\rm tanh}[\sigma (x+a)]  \ \
{\rm and} \ \ \chi \equiv {\rm tanh}[\sigma (x-a)]+1.
\end{equation}
The two-kink ansatz given by (\ref{2kinks}) can then be re-written as
\begin{equation}
\Phi(x) \approx {\Phi_- +\Phi_+ \over 2} + {\eta \over 2}Q^{(1)} f +
{\eta \over 2} Q^{(2)} (-1+\chi) \ .
\end{equation}
Initially, $\dot{\Phi}=0$. Also, within the range $x_1<x<x_2$, $\chi(x) \ll 1$ and
we can perform an expansion in $\chi$. To the leading order in $\chi$ we find:
\begin{eqnarray}
{dP \over dt} &\approx& \Big[ - {\eta^2\over 4} {\rm Tr}[(Q^{(1)})^2](f')^2
 - {\eta^2\over 2} {\rm Tr}[Q^{(1)}Q^{(2)}]f' \chi' \nonumber \\
 &+& \big[ V \big]_{\chi=0} +
\sum_a \big[{\partial V \over \partial \Phi^a}\big]_{\chi=0} \ \chi^a
\ \Big]_{x_1}^{x_2} \ ,
\end{eqnarray}
where functions $\chi^a$ are defined by
\begin{equation}
\chi^a \equiv {\eta \over 2} [Q^{(2)}]^a \chi
\end{equation}
and coefficients $[Q^{(2)}]^a$ are defined by
\begin{equation}
Q^{(2)} = \sum_a[Q^{(2)}]^a T^a \ ,
\end{equation}
where $T^a$ are the
$SU(N)$ generators normalized so that ${\rm Tr}[T^aT^b] = \delta_{ab}/2$. Using
the equations of motion gives
\begin{eqnarray}
{dP \over dt} &\approx& \Big[-{\eta^2 \over 2}{\rm Tr}[Q^{(1)}Q^{(2)}] f' \chi' +
\sum_a [{\Phi^a}'']_{\chi=0} \ \chi^a \ \Big]_{x_1}^{x_2} \nonumber \\
&=& {\eta^2 \over 2}{\rm Tr}[Q^{(1)}Q^{(2)}] \Big[-f' \chi' + f''\chi \Big]_{x_1}^{x_2} .
\end{eqnarray}
It can be shown that $\Big[ - f' \chi' + f''\chi \Big]_{x_1}^{x_2} < 0$ for
all $x_1$ and $x_2$ satisfying the constraints. Thus, the sign of $dP/dt$, and
the attractive or repulsive nature of the force, is determined by the sign
of ${\rm Tr}[Q^{(1)}Q^{(2)}]$.

Next we would like to express ${\rm Tr}[Q^{(1)} Q^{(2)}]$ in terms of parameters
$(q^{(1)},q^{(2)},q^{(3)},\epsilon_T^{(1)},\epsilon_T^{(2)})$ which define
the two-kink configuration. We write
\begin{eqnarray}
{\rm Tr}[Q^{(1)} Q^{(2)}] &=& {1 \over \eta^2}({\rm Tr}[\Phi_-\Phi_0]
+ {\rm Tr}[\Phi_0\Phi_+]
\nonumber \\ &-& {\rm Tr}[\Phi_-\Phi_+] - {\rm Tr}[\Phi_0\Phi_0]) \, .
\end{eqnarray}
Applying definitions (\ref{phiV}) and (\ref{phi+choices_N}) to boundary conditions
specified by $(\Phi_L,\Phi_R)$ we find:
\begin{eqnarray}
{1 \over \eta^2}{\rm Tr}[\Phi_L\Phi_R] = {2\epsilon_T \over N(N^2-1)}[(n+1-q)n^2
\nonumber \\ -2q^{(1)}(n+1)n
+ (n-q)(n+1)^2 ]
\nonumber \\ = {\epsilon_T\over 2} \left[1-q{(2n+1)\over n(n+1)}\right] \, .
\end{eqnarray}
Using the above expression and the fact that ${\rm Tr}[\Phi_0\Phi_0]=\eta^2/2$ we
find:
\begin{eqnarray}
{\rm Tr}[Q^{(1)} Q^{(2)}] = {1\over 2}(
\epsilon_T^{(1)}+\epsilon_T^{(2)}-
\epsilon_T^{(1)}\epsilon_T^{(2)} -1)
\nonumber \\ - (\epsilon_T^{(1)}q^{(1)}+\epsilon_T^{(2)}q^{(2)}
- \epsilon_T^{(1)}\epsilon_T^{(2)}q^{(3)}){(2n+1)\over 2n(n+1)} \, .
\label{nature}
\end{eqnarray}
One can see, for example, that in the case of a topological kink
interacting with a topological antikink ($\epsilon_T^{(1)}=-1$ and
$\epsilon_T^{(2)}=-1$) both, attraction and repulsion, are
possible depending on the choices of $q^{(1)}$, $q^{(2)}$ and
$q^{(3)}$. This is a novel feature, when compared to the classical
$\phi^4$ case \cite{Raj}, where the force between a kink and an
antikink is always attractive. An analogous situation is found in the case
of interactions between global $O(3)$ monopoles \cite{Leandros92}.
There, attraction or repulsion between a monopole and an antimonopole
is determined not by their topological charges but by the relative phase of their
field configurations.

Our derivation of the fact that the sign of ${\rm Tr}[Q^{(1)} Q^{(2)}]$
determines whether kinks will attract or repel was independent of a particular
form of $V(\Phi)$.
We did, however, rely on certain approximations in the derivation that may not be
valid for some extreme choices of $V(\Phi)$. Our numerical investigation of
kink interactions in $SU(5)\times Z_2$ with a general quartic
potential has always yielded an agreement between the sign of
${\rm Tr}[Q^{(1)} Q^{(2)}]$ and the overall sign of the
interaction potential.

In the next section we consider the $N=5$
case and study interactions between kinks and antikinks in more
detail.

\section{Kink interactions in $SU(5)\times Z_2$}
\label{interactions5}

Let us consider the
(1+1)-dimensional model of a scalar field
$\Phi$ transforming in the adjoint representation of  $SU(5)\times Z_2$.
We will take the potential to be
\begin{equation}
V(\Phi ) = - m^2 {\rm Tr}[ \Phi ^2 ]+ h ( {\rm Tr}[\Phi ^2  ])^2 +
      \lambda {\rm Tr}[\Phi ^4]  + V_0 \ ,
\label{potential}
\end{equation}
with parameters such that the vacuum expectation value (VEV) of
$\Phi$ breaks the symmetry spontaneously to $[SU(3)\times
SU(2)\times U(1)]/[Z_3\times Z_2]$. This happens in the parameter range
\begin{equation}
{h \over \lambda} > - {7\over {30}} \ , \ \ \lambda > 0 . \label{symmbreakparam}
\end{equation}
The VEV can be chosen to be
$\Phi_{V}=\eta \ {\rm diag}(2,2,2,-3,-3)/(2\sqrt{15})$, where
\begin{equation}
\eta \equiv {{m} \over {\sqrt{\lambda '}}} \label{eta}
\end{equation}
and
\begin{equation}
\lambda ' \equiv h + {7\over {30}} \lambda \ . \label{lambdaprime}
\end{equation}
The constant $V_0=m^2 \eta^2/4$ in Eq. (\ref{potential}) ensures
that $V(\Phi_{V})=0$.

We will study interactions between topological kinks and antikinks
($\epsilon_T^{(1)}=-1$, $\epsilon_T^{(2)}=-1$) with
all possible choices for indices $q^{(1)}$, $q^{(2)}$ and $q^{(3)}$,
as defined in Section \ref{kinks}. We will not consider interactions
between non-topological kinks nor interactions of non-topological kinks
with topological ones. Our initial configuration will be that of two
well-separated solitons moving towards each other.
From here on, we will label such configurations with three indices,
($q^{(1)}$, $q^{(2)}$, $q^{(3)}$), and it will be assumed that
$\epsilon_T^{(1)}=\epsilon_T^{(2)}=-1$.

\begin{table}[tbp]
\vskip 0.5 truecm
\begin{ruledtabular}
\begin{tabular}{|c|c|c|c|}
${\rm Tr}[Q^{(1)} Q^{(2)}]$ for & \ $q^{(3)}=0$ \  & \ $q^{(3)}=1$ \ & \ $q^{(3)}=2$\  \\ \hline
\ \ $q^{(1)}=0$, $q^{(2)}=0$ \ \ & \ $-2$ (A) \ & \ $\times$ \ & \ $\times$ \ \\ \hline
$q^{(1)}=0$, $q^{(2)}=1$  & $\times$  & \ $-{7 \over 6}$ (A) \ & $\times$ \\ \hline
$q^{(1)}=0$, $q^{(2)}=2$  & $\times$  & $\times$ & \ $-{1 \over 3}$ (A) \ \\ \hline
$q^{(1)}=1$, $q^{(2)}=1$  & $-{7 \over 6}$ (A)  & $-{3 \over 4}$ (A) & $-{1 \over 3}$ (A) \\ \hline
$q^{(1)}=1$, $q^{(2)}=2$  & $\times$  & $-{1 \over 3}$ (A) & $+{1 \over 12}$ (R) \\ \hline
$q^{(1)}=2$, $q^{(2)}=2$  & $-{1 \over 3}$ (A)  & $+{1 \over 12}$ (R) & $\times$
\end{tabular}
\end{ruledtabular}
\caption{${\rm Tr}[Q^{(1)} Q^{(2)}]$ evaluated for different choices
of the topological kink - topological antikink configuration in $SU(5)$.
(A) denotes attractions, (R) - repulsion and
$\times$ - means that the arrangement is impossible.}
\label{table}
\end{table}

In this model, only the $q=2$ topological kink solution is stable.
It also has the smallest energy of all kinks. However, we will not
restrict the analysis to $q=2$ kinks, as interactions between
all types of kinks antikinks could be potentially interesting.

We start by evaluating ${\rm Tr}[Q^{(1)} Q^{(2)}]$
for all possible choices of
the configuration of a kink and an antikink using Eq. (\ref{nature}). As we have shown in
Section \ref{interactionsN}, this whould tells us whether the two kinks will
attract or repel. The results are given
in Table \ref{table} and show that there are two possible combinations in which there is
a repulsive force between the solitons.

We would like to evaluate the interaction energy between different classes of kinks
and antikinks in $SU(5)\times Z_2$. The cases when analytical kink solutions are known are
the $q=0$ kink and, for the special value $h/\lambda= -3/20$,
the $q=2$ kink \cite{PogVac00,Vac01}.
The $q=0$ kink solution is the same as in the $\phi^4$ model, since in this case
the only non-zero component of $\Phi$ is the
one along $\Phi_{V}$, that is $\Phi=\phi(x) \Phi_V/\eta$, and the
potential takes the same form as in the $\phi^4$ theory:
\begin{equation}
V={\lambda'\over 4}(\phi^2-\eta^2)^2 \, .
\end{equation}
Therefore, the interaction potential between $q=0$ kinks and antikinks will be
identical to the one found in the $\phi^4$ model \cite{Moshir81}.

Let us next consider $q=2$ kinks and antikinks.
Consider a kink at $x=-a$ and an antikink at $x=a$, with $a>0$,
each moving with a velocity $v$ directed towards the origin. For values of
$a$ larger than the core sizes of two kinks the following ansatz is valid:
\begin{equation}
\Phi_{K\bar{K}}(x) = \Phi_K(x+a) + \Phi_{\bar{K}}(x-a) - \Phi_0,
\label{ansatz}
\end{equation}
where $ \Phi_0 = \Phi_K(+\infty)=\Phi_{\bar{K}}(-\infty)$ is the field in between
the two kinks. For $h=-3\lambda/20$ and $q=2$ the kink solution
is known \cite{PogVac00,Vac01}:
\begin{equation}
\Phi_{q=2}(x)={[1-\tanh(\sigma x)]\over 2}\Phi_- + {[1+\tanh(\sigma x)]\over 2}\Phi_+ \, ,
\end{equation}
where $\sigma = m/\sqrt{2}$.
The ansatz for the kink and antikink can then be written as
\begin{eqnarray}
\nonumber
\Phi_{K\bar{K}}(x) &=& {[1-F_K]\over 2}\Phi_- + {[1+F_K]\over 2}\Phi_0 \\
\nonumber
&+& {[1-F_{\bar{K}}]\over 2}\Phi_0 + {[1+F_{\bar{K}}]\over 2}\Phi_+  \\
&-& \Phi_0 \, ,
\label{ansatz1}
\end{eqnarray}
where $\Phi_-=\Phi_{K\bar{K}}(-\infty)$,
$\Phi_0=\Phi_{K\bar{K}}(0)$, $\Phi_+=\Phi_{K\bar{K}}(+\infty)$,
\begin{eqnarray}
F_K=\tanh[\sigma \gamma (x+a)], \\
F_{\bar{K}}=\tanh[\sigma \gamma (x-a)],
\end{eqnarray}
and $\gamma=1/\sqrt{1-v^2}$ is the Lorentz factor.

The total energy of the ansatz (\ref{ansatz1}) is a sum of the
gradient and the potential energies obtained by integrating
corresponding energy densities along the space direction $x$:
\begin{equation}
E = G + P \, ,
\end{equation}
where
\begin{equation}
\nonumber G = \int_{-\infty}^{+\infty} dx \ \textrm{Tr}(\partial_x
\Phi_{K\bar{K}})^2 \, ,
\label{kinenergy}
\end{equation}
\begin{eqnarray}
\nonumber P = \int_{-\infty}^{+\infty} dx \{ - m^2
\textrm{Tr}(\Phi_{K\bar{K}})^2 +
h\left(\textrm{Tr}(\Phi_{K\bar{K}})^2\right)^2 \\
\, \, \, + \lambda \textrm{Tr}(\Phi_{K\bar{K}})^4 + {{m^2
\eta^2} \over 4} \} \, \label{potenergy}
\end{eqnarray}

\begin{figure}
\vskip 0.5 truecm
\epsfxsize= 0.95\hsize\epsfbox{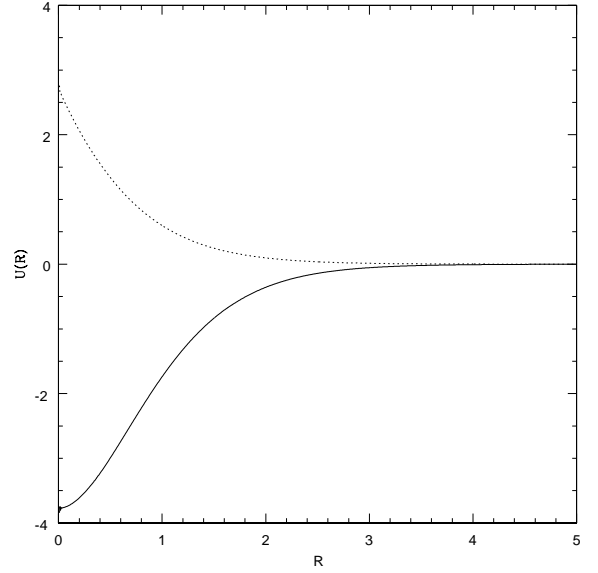}
\vskip 0.5 truecm
\caption{\label{fig:fig1} Interaction potentials $U_0$, given by Eq. (\protect \ref{energy3})
(solid line), and $U_1$, given by Eq. (\protect \ref{energy5}) (dotted line),
for $\gamma = m = \lambda = 1$.}
\end{figure}

At first let us consider a $q=2$ kink and a $q=2$ antikink such that
$\Phi_K(-\infty) = \Phi_{\bar{K}}(+\infty)$ (the $q^{(1)}=2$, $q^{(2)}=2$, $q^{(3)}=0$
case in Table \ref{table}). A possible set of boundary conditions
corresponding to this case is:
\begin{eqnarray}
\nonumber \Phi_- &=&
{\eta \over {2\sqrt{15}}} \textrm{diag}(2,2,2,-3,-3) \, , \\
\nonumber \Phi_0 &=&
{\eta \over {2\sqrt{15}}} \textrm{diag}(3,3,-2,-2,-2) \, , \\
\Phi_+ &=& {\eta \over {2\sqrt{15}}} \textrm{diag}(2,2,2,-3,-3) \,
. \label{bc1}
\end{eqnarray}
Using these boundary conditions in Eq.(\ref{ansatz1}) and
substituting latter into Eqns. (\ref{kinenergy}) and (\ref{potenergy}) gives
\begin{eqnarray}
\nonumber G &=& \int_{-\infty}^{+\infty} dX {m^3 \gamma \over
{\sqrt{2}\lambda}}\left(\partial_X F_K - \partial_X F_{\bar{K}}\right)^2 \, , \\
\nonumber P &=& \int_{-\infty}^{+\infty} dX {m^3 \over
{\sqrt{2}\lambda\gamma}} \left[ (F_K - F_{\bar{K}})^4 \right.\\
&-& \left. 4(F_K - F_{\bar{K}})^3 + 4(F_K - F_{\bar{K}})^2 \right] \, ,
\label{energy1}
\end{eqnarray}
where $X=\gamma m x/\sqrt{2}$ and we have taken $h/\lambda=-3/20$.
Evaluating integrals in Eq.(\ref{energy1}) and using $R=\sqrt{2} \gamma m a$
yields
\begin{eqnarray}
\nonumber G &=& {4\sqrt{2} m^3\gamma \over \lambda} \big[{1\over
3} + \frac{\sinh R - R\cosh R}{\sinh^3R}\big]\, , \\
\nonumber P &=& {4\sqrt{2} m^3 \over \lambda \gamma} \big[
{1\over3} + {1\over
\sinh^3R }\{ e^{-3R}({3\over 2}+R) \\
  &{}& \, \, \, \, \, \,  e^{-R}({7\over 2}R -
{1\over 2}) + e^{R}({1\over 2}R-1)\}\big]  \, . \label{energy2}
\end{eqnarray}
As expected, at $R=\infty$ the total energy is equal to the
sum of the two kink masses ($2\times 4\sqrt{2}m^3/(3\lambda)$)\cite{Vac01} divided by
the Lorentz factor. Subtracting $E(\infty)$ from $E(R)=G(R)+P(R)$ leaves only
the interaction part of the total energy:
\begin{eqnarray}
\nonumber U_0(R) = {4\sqrt{2} m^3 \gamma \over \lambda\sinh^3R}
\big( \sinh R - R\cosh R \\ \nonumber + {1\over \gamma^2}\big[
e^{-3R}({3\over 2}+R)
+e^{-R}({7\over 2}R-{1\over 2}) \\
+ e^{R}({1\over 2}R-1)\big] \big) \label{energy3}
\end{eqnarray}
The dependence of $U_0(R)$ on $R$ for $\gamma=1$ is shown as a solid line in
Fig. \ref{fig:fig1}. It clearly indicates an attraction
between the kink and the antikink.

The validity of the ansatz (\ref{ansatz}) cannot be justified for small
values of $R$. To test the analytical result, we have evaluated
the interaction energy numerically by explicitly integrating the full set of equations
of motion and evaluating the spatial integral over the gradient and potential
energy densities at each time step. The separation $R$ between the kinks was defined
as the distance between the maxima of their potential energy densities.
The shape of the found interaction potential was similar to the one given by
Eq. (\ref{energy3}), not only for $h/\lambda=-3/20$, but for all considered
values of the parameter: $-7/30 < h/\lambda < 100$. The interaction potential
for walls with different values of $\gamma>1$ was also qualitatively the same.

Next, let us consider the interaction of a $q=2$ kink and a $q=2$
antikink but with $\Phi_K(-\infty) \neq \Phi_{\bar{K}}(+\infty)$.
This would correspond to the ($q^{(1)}=2$, $q^{(2)}=2$, $q^{(3)}=1$)
case in Table \ref{table}. Keeping $\Phi_-$ and $\Phi_0$ the same as in Eq. (\ref{bc1})
we can choose $\Phi_+$ to be
\begin{eqnarray}
\Phi_+ &=& {\eta \over {2\sqrt{15}}} \textrm{diag}(2,2,-3,-3,2) \, .
\label{bc2}
\end{eqnarray}
This choice of boundary conditions leads to
\begin{eqnarray}
\nonumber G &=& \int_{-\infty}^{+\infty} dX {m^3 \gamma \over
{\sqrt{2}\lambda}}\left(2\partial_X F^2_K+\partial_x F_K \partial_X F_{\bar{K}}
+ 2\partial_xF^2_{\bar{K}} \right) \, , \\ \nonumber
P &=& \int_{-\infty}^{+\infty} dX {m^3 \over
{8 \sqrt{2}\lambda\gamma}} \left[ 8(F_K - F_{\bar{K}})^4 +8(F_K - F_{\bar{K}})^3\right. \\
\nonumber &-& 13(F_K - F_{\bar{K}})^2
- 30(F_K - F_{\bar{K}})(1-F_K F_{\bar{K}}) \\ \nonumber
&+& 40 F_K F_{\bar{K}}(F_K^2 + F_{\bar{K}}^2 - 2)
+ 10F_K F_{\bar{K}}(1- F_K F_{\bar{K}}) \\
&+& \left. 35(1-F_K^2 F_{\bar{K}}^2) \right] \, .
\label{energy4}
\end{eqnarray}
Performing the integration and subtracting $E(\infty)$ gives the
interaction potential between the two walls:
\begin{eqnarray}
\nonumber U_1(R) = {\sqrt{2} m^3 \gamma \over \lambda\sinh^3R}
\big( R\cosh R-\sinh R \\ \nonumber + {1\over
8\gamma^2}\big[ e^{-3R}(3+7R) -e^{-R}(11+13R) \\
 + e^{R}(8-4R) \big] \big) \label{energy5}
\end{eqnarray}
We find that in this case, the interaction potential is purely repulsive.
The dependence of $U_1(R)$ on $R$ for $\gamma=1$ is shown as a dashed line in
Fig. \ref{fig:fig1}. The exact numerical evaluation of the interaction
energy of this kink-antikink system, using the method outlined above,
did not show qualitative deviations from $U_1(R)$ for all considered
values of $h/\lambda$ and $\gamma$.

The two configurations: ($q^{(1)}=2$, $q^{(2)}=2$, $q^{(3)}=0$)
and ($q^{(1)}=2$, $q^{(2)}=2$, $q^{(3)}=1$), as well as the
well-studied $\phi^4$-equivalent case ($q^{(1)}=0$, $q^{(2)}=0$,
$q^{(3)}=0$), are the only ones for which analytical kink
solutions are known. Other configurations from Table \ref{table}
were treated only numerically. For all considered choices of
parameters and velocities, we did not see any deviation from the
predictions for the attraction or repulsion given in Table \ref{table}.

\section{$SU(5)\times Z_2$ kink-antikink collisions}
\label{collisions}

In this section we will study kink-antikink collisions. Even in the ``simple''
case of a $\phi^4$ kink colliding with an antikink the variety of possible outcomes
is surprisingly rich \cite{old1,old2,old3,old4,old5,Moshir81,Campbell_etal83}.
Depending on the incident velocity, the two kinks can annihilate, or they
can bounce off each other and never meet again, or they can
form an intermediate bound state, namely, they can bounce off each other several times
before ultimately separating or annihilating. The dependence on the incident velocity is
rather non-trivial, as was found in \cite{old1,old2,old3,old4,old5,Moshir81}
and investigated in detail in \cite{Campbell_etal83}. Namely,
it was found that, over a relatively small range
of initial velocities, intervals of initial velocity for which kink and antikink
capture each other alternate with regions for which the interaction concludes with escape
to infinite separations. In \cite{Campbell_etal83}, this alternation phenomenon was
attributed to a nonlinear resonance between the orbital frequency of the bound
kink-antikink pair and the frequency of characteristic small oscillations of the
field localized at the moving kink and antikink centers.

We will not attempt a study of exact dependence of outcomes of
kink-antikink collisions on initial velocities. The reason is that
in $SU(5)\times Z_2$ there are too many possible combinations and
there is an additional parameter, $h/\lambda$, which can possibly
affect the stability of kink solutions and the outcome of
kink-antikink collisions. Instead, we will describe the outcomes
for each of the combinations listed in Table \ref{table} and
illustrate the most interesting ones\footnote{Several animated
kink collisions can be viewed at
http://theory.ic.ac.uk/$\sim$LEP/su5kinks.html .}.

In order to study the collisions, we need to integrate the field equations
of motion forward in time. Without loss of generality, we can choose the
initial kink-antikink configuration to be diagonal \cite{PogVac01}.
Since equations of motion preserve the diagonal form, one only needs to
consider the evolution of four functions: $a(x,t)$, $b(x,t)$, $c(x,t)$ and $d(x,t)$,
defined as:
\begin{equation}
\Phi(x,t)=a(x,t)\lambda_3+b(x,t)\lambda_8+c(x,t)\tau_3+d(x,t)Y \ ,
\label{components}
\end{equation}
where $\lambda_3$, $\lambda_8$, $\tau_3$ and $Y$ are the
diagonal generators of SU(5):
\begin{eqnarray}
\lambda_3&=&\frac{1}{2} {\rm diag}(1,-1,0,0,0) \ , \nonumber \\
\lambda_8&=&\frac{1}{2\sqrt{3}} {\rm diag}(1,1,-2,0,0),
\nonumber \\
\tau_3&=&\frac{1}{2} {\rm diag}(0,0,0,1,-1) \ , \nonumber \\
Y&=&\frac{1}{2\sqrt{15}} {\rm diag}(2,2,2,-3,-3) \ .
\label{ymatrix}
\end{eqnarray}

\begin{figure}[tbp]
\epsfxsize= 0.9\hsize\epsfbox{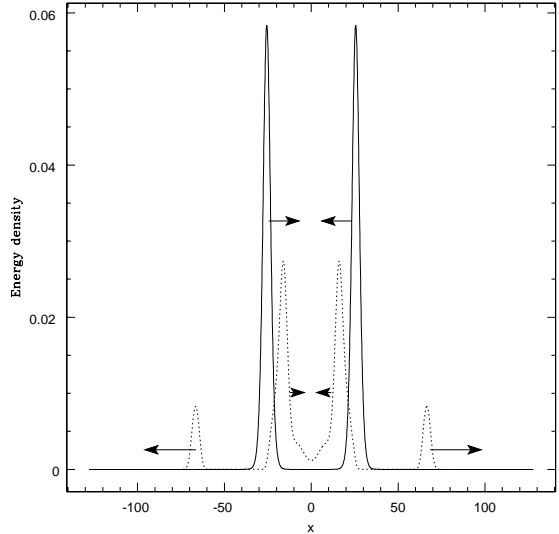} \caption{\label{fig:q000-en}
Kink-antikink collision in the ($q^{(1)}=0$, $q^{(2)}=0$,
$q^{(3)}=0$) case, in the parameter range when $q=0$ kink is
unstable. The solid line shows the initial energy density profile
and the dotted line - the final. The right moving $q=0$ kink
collapsed into a $q=2$ kink going to the left and the remainder
$q=1$ non-topological ($\epsilon_T=1$) kink moving to the right.
The originally left moving $q=0$ antikink splits into a right
moving $q=2$ antikink and a left moving $q=1$ non-topological
kink. The arrows show directions and approximate
relative magnitudes of kink velocities.}
\end{figure}
\begin{figure}[tbp]
\epsfxsize= 0.9\hsize\epsfbox{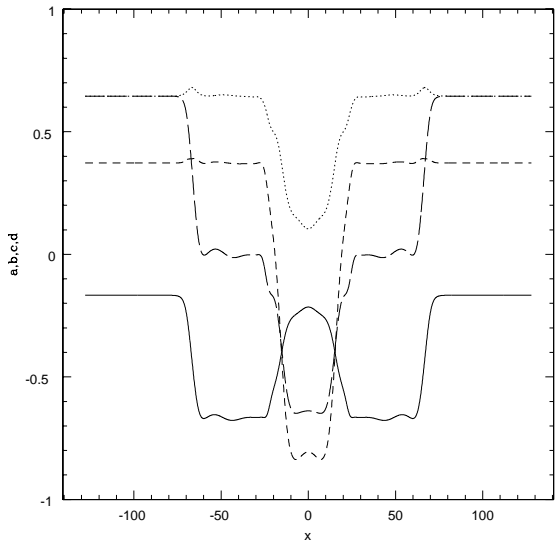} \caption{\label{fig:q000-fields}
Functions $a(x,t)$ (dotted line), $b(x,t)$ (short dash line), $c(x,t)$ (long dash line)
and $d(x,t)$ (solid line),
defined in Eq. (\protect \ref{components}) at the same final snapshot as in
Fig. \protect \ref{fig:q000-en}.}
\end{figure}
As mentioned in Section \ref{interactionsN}, the $q=0$ wall
is identical to the kink of the simplest $\phi^4$-model. The $q=0$
solution, however, is known to be locally unstable against perturbations
along diagonal components of $\Phi$ in the parameter range
$h/\lambda > -3/20$ \cite{DvaLiuVac98}. The outcome of the $q=0$ kink-antikink collision
will, therefore, depend on $h/\lambda$ as well as on $v_{initial}$.
In Fig. \ref{fig:q000-en} and Fig. \ref{fig:q000-fields} we illustrate the collision for
$v_{initial}=0.05$ and $h/\lambda=0$. Just before the collision, both the kink and the
antikink collapse. More detailed analysis of functions
$a$, $b$, $c$ and $d$ in Fig. \ref{fig:q000-fields}  reveals that the $q=0$ kink, initially
moving to the right, has collapsed into a $q=2$ kink, moving to the left, and
the remainder $q=1$ non-topological ($\epsilon_T=1$) kink, moving to the right.
The originally left moving $q=0$ antikink has split into a right moving $q=2$ antikink
and a left moving $q=1$ non-topological kink. The $q=2$ kinks will separate to infinities,
while the fate of $q=1$ non-topological kinks needs more explanation.
As was found in \cite{PogVac01}, the
$q=1$ non-topological kink is unstable against a collapse into a pair of $q=2$ kinks.
Therefore, depending on the values of $v_{initial}$ and $h/\lambda$, the non-topological
$q=1$ kinks (in the center of Fig. \ref{fig:q000-en}) can either immediately decay into
radiation, or they can split into pairs of $q=2$ kinks moving away from each other.
We found that the latter was the case for the choice of parameters corresponding to
Fig. \ref{fig:q000-en} and Fig. \ref{fig:q000-fields}.
For $-3/20 < h/\lambda < -3/70$, when $q=0$ kinks are locally stable, outcomes of their
collisions are the same as in the case of kinks and antikinks in
$\phi^4$ model, the case studied extensively
in \cite{old1,old2,old3,old4,old5,Moshir81,Sugiyama79,Campbell_etal83}.
\begin{figure}[tbp]
\epsfxsize= 0.9\hsize\epsfbox{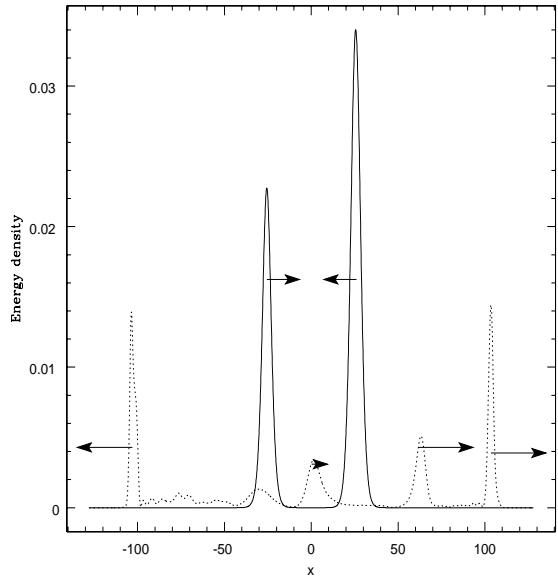}
\caption{\label{fig:q011} Kink-antikink collision in
the ($q^{(1)}=1$, $q^{(2)}=0$, $q^{(3)}=1$) case. The solid line shows the initial
energy density profile and the dotted line - the final. After the
collision, there are four $q=2$ kinks arranged so that the left most and
the right most pairs of kinks form ($q^{(1)}=2$, $q^{(2)}=2$, $q^{(3)}=1$)
combinations and the two inner kinks form a ($q^{(1)}=2$, $q^{(2)}=2$, $q^{(3)}=0$)
combination. The rest of the original energy is being radiated away.
The arrows show directions and
approximate relative magnitudes of kink velocities.}
\end{figure}

Fig. \ref{fig:q011} illustrates a collision of a $q=1$ kink
(initially on the left) with a $q=0$ antikink (initially on the
right). The parameters are $v_{initial}=0.2$, $\eta=1$,
$\lambda=0.5$ and $h/\lambda = -0.1$. The $q=0$ kink is unstable
for these parameters. The mass of the $q=0$ kink is
$2\sqrt{2}m^3/\lambda'$ \cite{Raj} and, for our choice of
parameters, is equal to $0.243$. The $q=1$ kink mass can only be
found numerically and is $0.150$. For comparison, the mass of the
$q=2$ kink with these parameters would be $0.033$ - almost one
fifth the mass of the $q=1$ kink. We find that, during the
collision, the $q=0$ kink collapses into a $q=2$ kink traveling
to the right and a $q=1$ non-topological kink traveling to the
left. At the same time, the original $q=1$ kink collapses into
three $q=2$ kinks, with outer kinks moving away from each other.
The final configuration is that of four $q=2$ kinks arranged so
that the two left most kinks and the two right most kinks form
($q^{(1)}=2$, $q^{(2)}=2$, $q^{(3)}=1$) combinations, while the
two inner kinks form a ($q^{(1)}=2$, $q^{(2)}=2$, $q^{(3)}=0$)
combination. The rest of the original energy is radiated away.

\begin{figure}[tbp]
\epsfxsize= 0.9\hsize\epsfbox{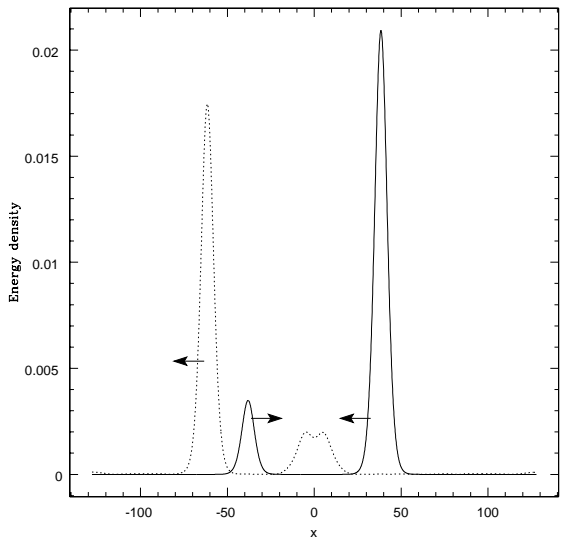}
\caption{\label{fig:q022} Kink-antikink collision in
the ($q^{(1)}=0$, $q^{(2)}=2$, $q^{(3)}=2$) case. Resulting configuration (dotted line)
is that of a non-topological $q=1$ kink moving to the left, while the $q=2$
kink has captured its ``mirror image'' (the $q=2$ antikink) originally contained in the
incident $q=1$ wall.}
\end{figure}

In Fig. \ref{fig:q022} the initial configuration is
($q^{(1)}=0$, $q^{(2)}=2$, $q^{(3)}=2$).
In this case, the $q=0$ antikink (initially on the right) splits into a
$q=2$ topological antikink, which starts interacting with the $q=2$ kink,
and a $q=1$ non-topological kink, which keeps propagating to the left unperturbed.
Another way to describe this interaction is that
the $q=2$ wall has met its $q=2$ ``reflection'' in
the $q=0$ wall and formed a complex with it, while the
remainder wall is radiated away.

The ($q^{(1)}=1$, $q^{(2)}=1$, $q^{(3)}=0$) case has essentially the same set of
possible outcomes as the ($q^{(1)}=0$, $q^{(2)}=0$, $q^{(3)}=0$) and will not be
considered here.

\begin{figure}[tbp]
\epsfxsize= 0.9\hsize\epsfbox{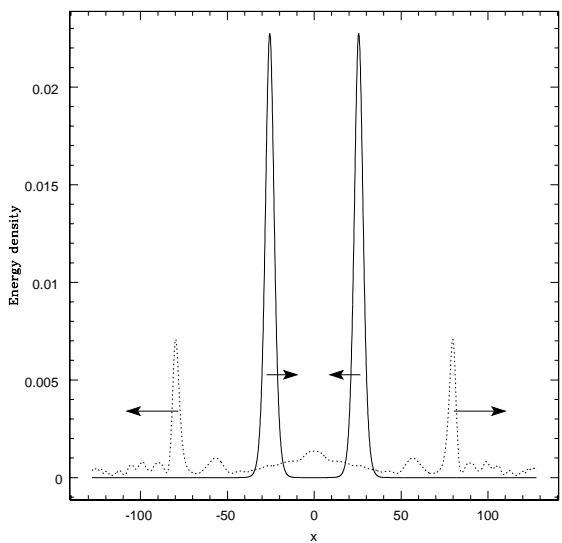}
\caption{\label{fig:q111} The ($q^{(1)}=1$, $q^{(2)}=1$, $q^{(3)}=1$) case with
$h/\lambda = -0.1$ and $v_{initial}=0.2$. The final configuration is that of two
$q=2$ kinks, arranged in a
($q^{(1)}=2$, $q^{(2)}=2$, $q^{(3)}=1$) combination, moving away from each other.}
\end{figure}

Fig. \ref{fig:q111} illustrates the outcome of the kink-antikink collision
with the initial ($q^{(1)}=1$, $q^{(2)}=1$, $q^{(3)}=1$) configuration. The
parameters were chosen to be $h/\lambda = -0.1$ and $v_{initial}=0.2$.
The final configuration is that of two $q=2$ kinks, arranged in a
($q^{(1)}=2$, $q^{(2)}=2$, $q^{(3)}=1$) combination, moving away from each other.

\begin{figure}[tbp]
\epsfxsize= 0.9\hsize\epsfbox{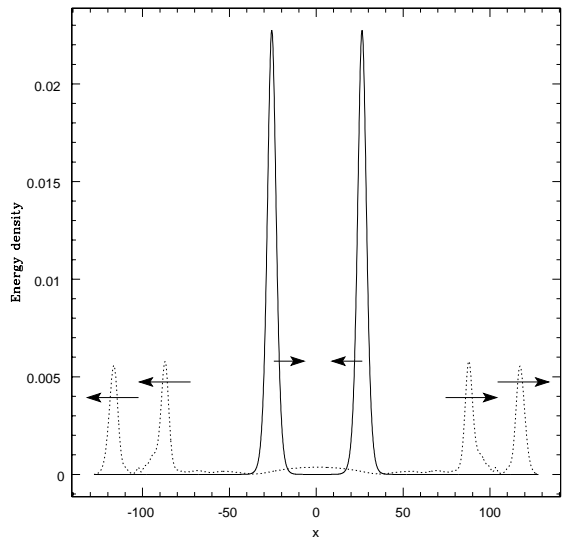}
\caption{\label{fig:q112} The ($q^{(1)}=1$, $q^{(2)}=1$, $q^{(3)}=2$) case with
$h/\lambda = -0.1$ and $v_{initial}=0.2$. The final configuration is that two ($q^{(1)}=2$, $q^{(2)}=2$, $q^{(3)}=1$)
combinations moving away from each other.}
\end{figure}

The outcome of a ($q^{(1)}=1$, $q^{(2)}=1$, $q^{(3)}=2$) collision with
$h/\lambda = -0.1$ and $v_{initial}=0.2$ is shown in Fig. \ref{fig:q112}.
The final configuration is that two ($q^{(1)}=2$, $q^{(2)}=2$, $q^{(3)}=1$)
combinations moving away from each other.

\begin{figure}[tbp]
\epsfxsize= 0.9\hsize\epsfbox{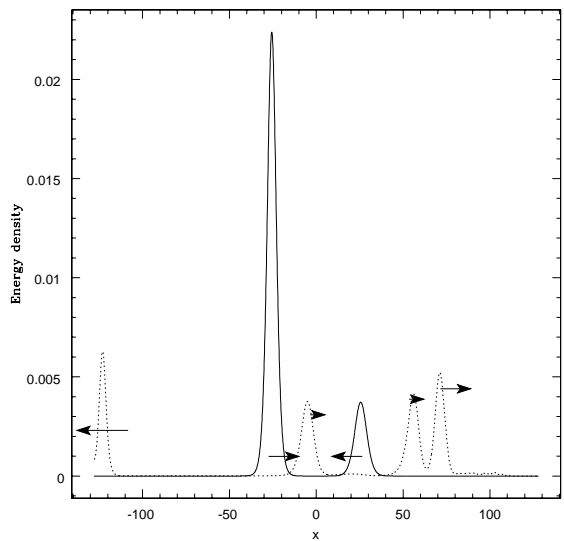}
\caption{\label{fig:q121} The ($q^{(1)}=1$, $q^{(2)}=2$, $q^{(3)}=1$) case with
$h/\lambda = -0.1$ and $v_{initial}=0.1$. There are four $q=2$ kinks left as a result
of the collision.}
\end{figure}

The ($q^{(1)}=1$, $q^{(2)}=2$, $q^{(3)}=1$) case with $h/\lambda =
-0.1$ and $v_{initial}=0.1$ is illustrated in Fig. \ref{fig:q121}.
Just before the collision, the $q=1$ kink (originally on the left)
collapses into three $q=2$ kinks, with two outer kinks having
large kinetic energies. Thus, an intermediate configuration is
that of four $q=2$ kinks, including the initial one (originally on
the right). The two right most $q=2$ kinks are in a ($q^{(1)}=2$,
$q^{(2)}=2$, $q^{(3)}=0$) combination and, depending on the
initial velocity, may annihilate or chase each other forever. The
two left most walls are in a ($q^{(1)}=2$, $q^{(2)}=2$,
$q^{(3)}=1$) arrangement.

A $q^{(1)}=1$ kink and a $q^{(2)}=2$ antikink arranged in a
($q^{(1)}=1$, $q^{(2)}=2$, $q^{(3)}=2$) combination repel.
For $h/\lambda = -0.1$ and $v_{initial}=0.15$ they scatter elastically.
The lighter $q=2$ kink (initially on the right) bounces off
the heavier $q=1$ kink and slows it down.
\begin{figure}[tbp]
\epsfxsize= 0.9\hsize\epsfbox{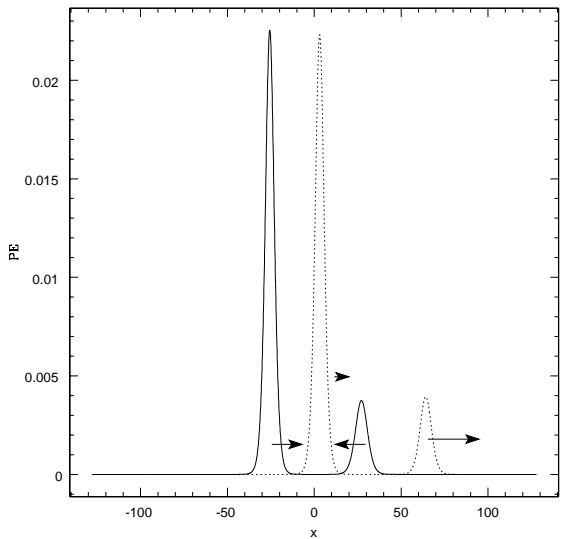}
\caption{\label{fig:q122} The elastic scattering in the
($q^{(1)}=1$, $q^{(2)}=2$, $q^{(3)}=2$) case with $h/\lambda = -0.1$ and $v_{initial}=0.1$.}
\end{figure}

The remaining two combinations from Table \ref{table} are
($q^{(1)}=2$, $q^{(2)}=2$, $q^{(3)}=0$) and ($q^{(1)}=2$, $q^{(2)}=2$, $q^{(3)}=1$).
Since these are the only two initial combinations that involve stable
kinks and antikinks, they would be the most important ones if one studied
evolution of domain wall networks after the formation. However, these two
are also the most ``uninteresting'' combinations from the novelty point of veiw.
In the case of ($q^{(1)}=2$, $q^{(2)}=2$, $q^{(3)}=0$), the dynamics and possible
outcomes are qualitatively identical to the case of kink-antikink collision
in the simple $\phi^4$ model.
Namely, we observe a dependence on the incident velocity similar to that found in
Refs.~\cite{old1,old2,old3,old4,old5,Moshir81,Campbell_etal83}.
The difference is that in the case of $SU(5)$ the value of the incident velocity
leading to a given outcome depends on the parameter $h/\lambda$ of the potential given in
Eq.~(\ref{potential}).

In the ($q^{(1)}=2$, $q^{(2)}=2$, $q^{(3)}=1$) case,
the kink and antikink repel and simply bounce off each other elastically.

\section{Summary and discussion}
\label{conclude}
As we have illustrated in the previous sections,
kink-antikink interactions in $SU(N)\times Z_2$ can be of different types
with many possible outcomes depending on the choice of the parameter values.

In order to see how a pair of kinks will interact, one generally needs
to specify the potential $V(\Phi)$ and evaluate the interaction energy between the
two kinks. However, we have shown that to a very good approximation the nature of
the interaction can be determined by evaluating ${\rm Tr}[Q^{(1)}Q^{(2)}]$,
where $Q^{(1)}$ and $Q^{(2)}$ are the charges of the two kinks defined by
Eq. (\ref{charge}). In particular, we have shown that kinks and antikinks may
attract or repel depending on their relative orientation in the internal space.
This is similar to interactions between global $O(3)$ monopoles, where the relative phase,
not the topological charges, is what determines the nature of the
interaction \cite{Leandros92}.

In our study of kink-antikink collisions in $SU(5)\times Z_2$ we have seen
a general tendency for larger mass topological kinks to split into fundamental kinks
of the theory, such as the $q=2$ kink. This suggests that kink-antikink interactions in
$SU(N)\times Z_2$ can be described in terms of interactions between
fundamental ({\it i.e.} globally stable) $q=(N-1)/2$ kinks. Namely, given
the configuration of two kinks, one would look for the least energy configuration
of $q=n$ kinks that would have the same global topology as the original
pair of kinks. The outcome of the interaction would then be reduced to interactions
between attractive or repulsive pairs of $q=n$ kinks.

We have not investigated the detailed dependence of the dynamics of
kink-antikink collisions on the initial velocities. Part of the
reason is that the outcome strongly depends on the
stability properties of each of the solitons and our numerical methods
do not give us the possibility of properly accounting for all instabilities
that can occur in the model. Each kink, except for the
fundamental one, is unstable in a different way, namely, along a different
direction in the internal space of SU(5), and also depends on the particular choice
of the potential. It is possible that a more detailed study would reveal a
connection with earlier work on Q balls \cite{Axenides} and global $U(1)$
strings \cite{Shellard}, where is was observed that at very high collision velocities
fragmentation of solitons is generally suppressed. Such a study could be
a subject of future work.

From the equations of motion it follows that if the Higgs field
components along non-diagonal generators of $SU(5)$ were zero at
the initial time, they would remain zero at all times, which was
the case in our simulations. However, except for the $q=2$
topological kinks, kink solutions can be unstable against
perturbations along non-diagonal generators of $SU(5)$
\cite{PogVac01}. In a realistic domain wall formation scenario one
would have to allow for all components of the Higgs to be excited
and only stable walls would survive. Depending on the rate of the
phase transition, temporary formation of unstable kinks will or
will not be relevant. Nevertheless, stable and unstable kinks are
solutions of the classical field equations and their interactions
may become important whenever an exact or approximate $SU(N)\times
Z_2$ symmetry is present in a theory.

\acknowledgments

I would like to thank Tanmay Vachaspati for several important
insights and for commenting on the earlier draft of the
manuscript. I am grateful to N.~Antunes, J.~Kalkkinen and T.~W.~B.~Kibble for
useful discussions. Conversations with participants of the ESF COSLAB
Programme workshop at Imperial College in July, 2001, and
especially with F.~A.~Bais and L.~Perivolaropoulos, are acknowledged.
This work was supported by PPARC.

\end{document}